\documentclass[doublecol]{epl2} 
\usepackage{graphicx}
\usepackage{amsmath}
\usepackage{revsymb}
\usepackage{bm}

\title{Spin-wave spectra of a kagome stripe}
\author{A. Donkov\inst{1} \and A. V. Chubukov\inst{2}}
\shortauthor{A. Donkov and A. V. Chubukov}

\institute{                    
  \inst{1} Max-Planck-Institut f\"ur Physik Komplexer Systeme, D-01187 Dresden, Germany \\
  \inst{2} Department of Physics, University of Wisconsin, Madison, WI 53706, USA}
\abstract{We study ground state degeneracy and spin-wave excitations in
 a 1D version of a Kagome antiferromagnet -- a
 Heisenberg antiferromagnet on a  Kagome stripe.
We show that for nearest-neighbor interaction, the classical ground state is infinitely degenerate. For any spin configuration from the degenerate set,
 the classical spin-wave spectrum contains, in addition to Goldstone modes,
 a branch  of zero energy excitations, and a zero mode in
 another branch. We demonstrate that the interactions 
beyond nearest neighbors lift the degeneracy, eliminate a zero mode, and give  a
 finite dispersion to formerly zero-energy branch, 
leaving only Goldstone modes as zero-energy excitations. }

\begin{document}

\maketitle

In the last few years, there has been a revival of interest in the 
 studies of frustrated spin systems. One of the most intriguing aspects of 
spin frustration is the existence, in many cases, of extra zero modes in the excitation spectrum, in addition to Goldstone modes related to the breaking of a 
 continuous symmetry. These zero modes are often associated with the 
local degeneracy of a classical ground state of a frustrated system
 with short-range interactions between nearest-neighbors, and 
 are lifted either by fluctuations, thermal or quantum, or by 
 longer-range interactions which include second, third, etc neighbors~\cite{ofd}.

The most prominent example of a system with extra zero modes is a much studied 
2D nearest-neighbor antiferromagnet on a Kagome lattice~\cite{ZengElser90,Sachdev,Harrisetal92,Chubukov92}.  The Kagome lattice
 consists of corner-sharing hexagons, and can be obtained from a triangular lattice by removing a
 quarter of the spins.
 The classical spectrum of a Kagome antiferromagnet contains the whole branch of zero-energy excitations, 
associated with the local degeneracy of a classical ground state with respect to rotations of spins belonging to a particular hexagon.  Quantum and thermal fluctuations remove the degeneracy and select a particular ``$\sqrt{3} \times \sqrt{3}$'' configuration, same as in a triangular antiferromagnet~\cite{Sachdev,Chubukov92,Zhitomirsky:2002PRL}.  The same selection, also accompanied by the lifting of zero modes, can be also achieved by adding
 interactions between next-nearest neighbors~\cite{Harrisetal92}.
 In this later case, the lifting of the degeneracy is a natural consequence of the fact that next-nearest-neighbor
 interaction connects spins belonging to different hexagons
 and adds an energy cost to local rotations. 

Most of recent work on Kagome-type systems was devoted to 3D version of a Kagome antiferromagnet, which is an antiferromagnet of the pyrochlore lattice~\cite{pyro}.
Less attention was given to an ``opposite'' 1D version, which is an antiferromagnet on a Kagome stripe. This is a three-chain structure, consisting of 
 of top/bottom sharing 
hexagons (Fig. 1b).  Like its 2D parent, a
 Kagome-stripe antiferromagnet can be obtained from a three-chain triangular antiferromagnet by removing 1/6 of the spins (Fig. 1a). 
 The existing analytical\cite{Azariaetal98,Azariareply} and 
numerical\cite{PatiSingh99,WhiteSingh00} works
 focused on $S=1/2$ and primarily 
addressed the issue of a spin-disordered state with gapped spin-triplet excitations, and gapless spin-singlet excitations. Closely related three-spin
ladder system have been investigated in \cite{Waldtmannetal00} by
exact diagonalization and DMRG techniques. The
 studies of Kagome stripes in high magnetic fields have been performed  
in \cite{DerzhkoRichter06,Schnacketal06,Schmidtetal06}.
\begin{figure}[htb!]
\centering%
\includegraphics{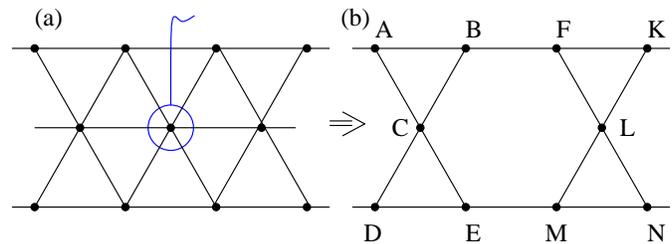} 
\caption{\label{fig:lattice} 
(a) A triangular three-chain stripe. The Kagome stripe (b) is obtained by
 taking out the spin in the 
middle.}
\end{figure}

In this communication, we analyze the properties of a Heisenberg 
Kagome stripe with a large spin $S\gg1$.  We will not discuss the destruction of long-range magnetic order 
 by 1D fluctuations, which at $T=0$ 
occurs only at exponentially small energies, but rather focus on the issue of 
 the lifting of the ground-state degeneracy and corresponding zero modes 
by the interaction beyond nearest neighbors. 
We show that there are two {\it different}
  ground state degeneracies in a Kagome 
stripe, besides a conventional global degeneracy which is broken by the 
long-range order and gives rise to Goldstone modes. One is
 a truly local degeneracy, which gives rise to a whole 
branch of zero-energy excitations. Another is an extra  global degeneracy with gives rise to an extra zero-mode at a particular momentum $k=0$.  We demonstrate that the local degeneracy 
 is lifted by the interactions between spins in the middle chain (formally, third-neighbor interaction), while the 
extra global  degeneracy is lifted by a second-neighbor 
interaction in the direction perpendicular to the direction of the chains.

\begin{figure}[htb!]
\centering%
\includegraphics{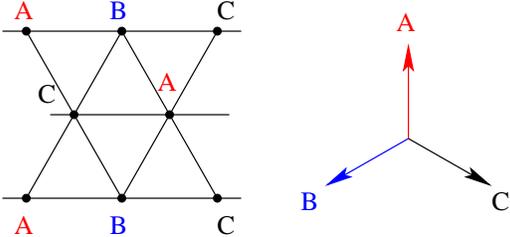} 
\caption{\label{fig:nonfrustr} 120$^\circ$ spin ordering on a triangular stripe. The directions of the spin quantization axes are labeled by $A,B,C.$
The ordering of the whole system is uniquely determined by the ABC 
ordering in a given triangle. }
\end{figure}

The frustrated nature of the Kagome stripe can be most easily understood by comparing it with the triangular stripe in Fig.~\ref{fig:lattice}.  In both cases, the lowest energy of a particular triangle of spins is achieved by placing the spins $120^\circ$ apart.  For a triangular stripe with nearest-neighbor interaction, the $120^\circ$ ordering of a particular triangle 
 uniquely determines the ordering  of the full system (see Fig.~\ref{fig:nonfrustr}). From this perspective, the triangular stripe of Heisenberg spins is non-frustrated.  
For a Kagome stripe  with nearest-neighbor interaction, the $120^\circ$ ordering of a particular triangle does not 
specify the global order by two reasons. First, the spins D and E in the triangle DEC, connected to the ABC triangle by the spin C in the middle chain (see Fig. 1b),  can freely rotate around the quantization 
axis of the middle-chain spin C (such rotation is impossible in a triangular antiferromagnet as there the spins D and E are also connected to the neighboring spins in the middle chain, whose directions are fixed by the ordering in ABC triangle.  Second,  the five spins in the next set of two triangles FKL and LMN, 
sharing the spin L in the middle chain,
 are connected to the triangle ABC only via antiferromagnetic interaction between nearest neighbors B and F along the upper chain.  This sets the direction
 of the spin F, but other four spins can rotate in various ways preserving
 a $120^\circ$ ordering within  triangles FKL and LMN. This is clearly a 
 frustrated system.
 
\begin{figure}[t!]
\centering%
\includegraphics{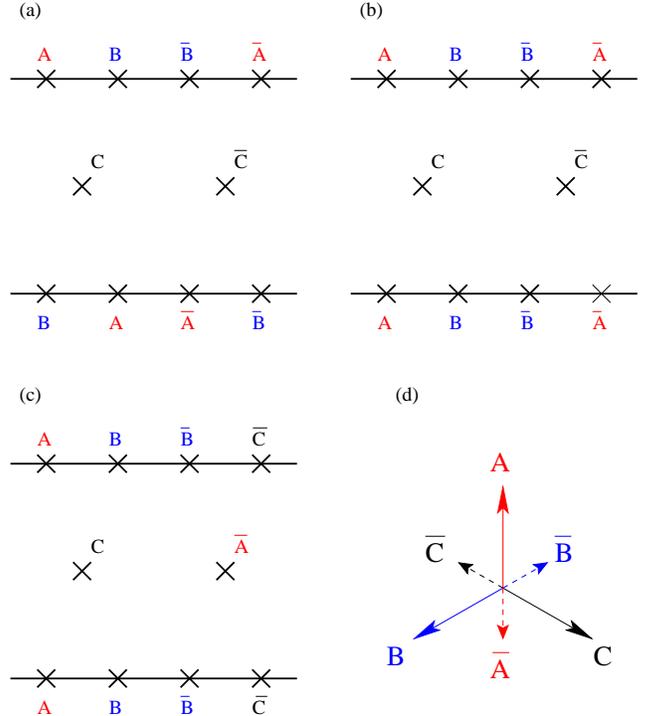} 
\caption{\label{fig:gndchoice} Different ground state configurations on a Kagome stripe. Each preserves a $120^\circ$ ordering within each spin triangle, and antiparallel orientation of the coupled spins from different triangles.
 The directions $A,B,C$ and their opposites  ${\bar A}, {\bar B}, {\bar C}$ are shown in panel (d). The configuration (b) is the 1D version of the $\sqrt{3} \times \sqrt{3}$  configuration of a 2D  Kagome antiferromagnet. It is stabilized 
 by the interactions between second and third neighbors. For 
 nearest neighbor interaction only, the spins in 
the configuration 2b can be moved without energy cost, as  shown in Figs.~\ref{fig:J2lift} and \ref{fig:J3lift}. 
 }
\end{figure}

\begin{figure}[t!]
\centering%
\includegraphics{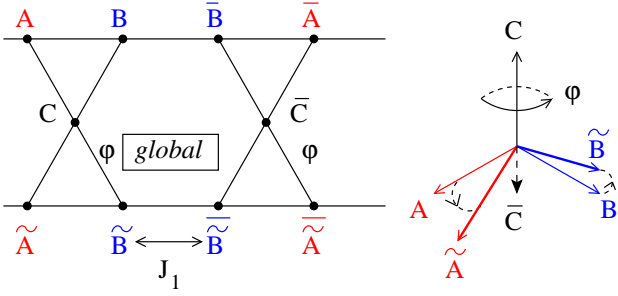} 
\caption{\label{fig:J2lift} The ``global'' degeneracy of the 
 $\sqrt{3} \times \sqrt{3}$ spin configuration. 
Spins along the lower chain can rotate around the common quantization axis of 
 middle-chain spins $C$ and ${\bar C}$, while the spins along the upper chain remain intact. Since the spins along the lower chain belonging to different 
 sets of top/bottom sharing triangles must remain antiparallel to each other, this rotation simultaneously involves all spins from the lower chain, and is therefore a global rotation. This degeneracy gives rise to a zero mode in the spin-wave spectrum.}
\end{figure}

In Fig. \ref{fig:gndchoice} we show three different ground state configurations of a Kagome stripe, each preserves a $120^\circ$ ordering within each spin triangle, and antiparallel orientation of the coupled spins from different triangles. For nearest-neighbor interaction, the spin-wave spectra of all
 these configurations are equivalent. The equivalence is broken, however, 
 once we include the interactions between next-nearest neighbors.  For definiteness, we focus on the configuration shown in Fig. \ref{fig:gndchoice}b -- this is 1D version of the $\sqrt{3} \times \sqrt{3}$ configuration of a 2D Kagome antiferromagnet.  
For the configuration of Fig. \ref{fig:gndchoice}b, the
  existence of two different ground state degeneracies can be easily visualized. Indeed, the spins in the middle chain have the same quantization axis  
(the axis is understood here as a director rather than a vector). One can then 
 rotate the spins in e.g., lower chain around this axis, leaving the spins in the upper chain intact, see Fig.~\ref{fig:J2lift}.  As the spins along the lower chain, belonging to neighboring triangles, interact antiferromagnetically and must remain antiparallel, the rotation must simultaneously involve all spins in the lower chain, and is therefore a {\it global} rotation. The degeneracy associated with such rotation is obviously lifted by an interaction $J_2$ between a spin in the lower chain and 
 a spin on top of it in the upper chain.  A ferromagnetic $J_2$
 makes
 the configuration of Fig. \ref{fig:gndchoice}b stable.   

\begin{figure}[t!]
\centering%
\includegraphics{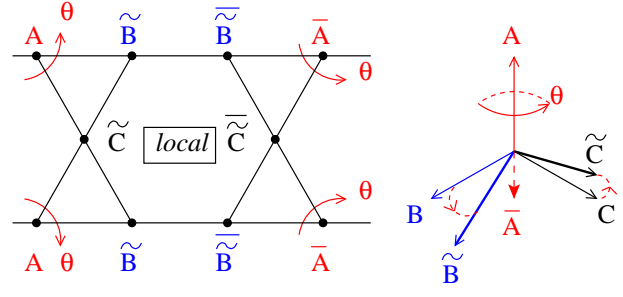} 
\caption{\label{fig:J3lift} The ``local'' 
 degeneracy of the 
 $\sqrt{3} \times \sqrt{3}$ spin configuration. 
Spins along the hexagon ${\tilde B}, ~{\bar{\tilde B}},~{\bar{\tilde C}},~
{\bar{\tilde B}},~{\tilde B}, {\bar{\tilde C}}$ can rotate around the
around the common quantization axis of the ``corner'' spins $A$ and ${\bar A}$,  preserving the $120^\circ$ orientations of spins in any triangle.
  This degeneracy is {\it local}
 as it only involve spins inside a particular hexagon.
The local degeneracy gives rise to zero-energy branch of spin-wave excitations.}
\end{figure}

  Another degeneracy is associated 
 with the rotation of the six spins of a  hexagon $B, {\bar B}, {\bar C}, {\bar B}, B, C$ around the common quantization axis of the ``corner'' spins $A$ and ${\bar A}$, such that $120^\circ$ orientations of spins in any triangle is preserved, see Fig.~\ref{fig:J3lift}.  This degeneracy is {\it local}
 as it only involve spins inside a particular hexagon, while the spins and $A$ and ${\bar A}$, through which  a given hexagon is connected to other hexagons,
 remain intact.  The  local degeneracy is lifted by the 
interaction $J_3$ between the spins in the middle chain, 
as the angle between these spins obviously changes in the process of a local rotation. An antiferromagnetic $J_3$ obviously favors antiparallel orientation of the spins $C$ and ${\bar C}$ and therefore stabilizes the configuration in Fig. \ref{fig:gndchoice}b.  

\begin{figure}[t!]
\centering%
\includegraphics{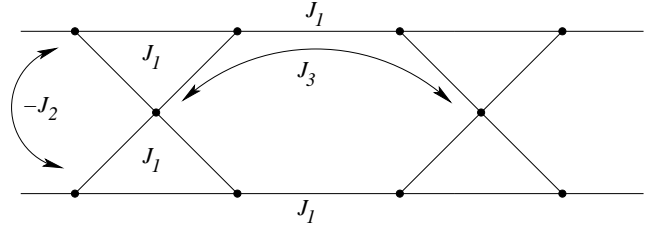} 
\caption{\label{fig:latticeJ1J2J3}Kagome stripe with antiferromagnetic nearest neighbor  and third-neighbor interactions,  $J_1$ and $J_3$, respectively, and 
a  ferromagnetic second-neighbor exchange interaction  $J_2$. 
The Hamiltonian for this model is given by Eq.~(\ref{eq:hamiltonian}).}
\end{figure}

To this end, we consider a $J_1-J_2-J_3$ model with Heisenberg interaction
 within a Kagome stripe ($J_1$), in between two chains ($J_2$), and 
 between the spins in the middle chain $(J_3)$ (see Fig.\ref{fig:latticeJ1J2J3}). The corresponding Hamiltonian is
\begin{equation}
H = J_1 \sum_{<i,i'>} {\textbf S}_i {\textbf S}_{i'} - J_2 
\sum_{i} {\textbf S}_{i,l} {\textbf S}_{i,u} + J_3 \sum {\textbf S}_{i,m}{\textbf S}_{j,m}.
\label{eq:hamiltonian}
\end{equation}
where indices $i,i'$ denote the spins belonging to the same set of two 
top/bottom sharing triangles, $i,j$ denote states belonging to different triangles, and $l,u,m$ denote spins from lower, upper, and middle chains, respectively.
 We obtain the spin-wave  spectrum of this Hamiltonian and show explicitly that 
 the interactions $J_2$ and $J_3$ lift the zero modes leaving only three 
Goldstone 
 zero-energy excitations, associated with $\sqrt{3} \times \sqrt{3}$ ordering.
These Goldstone modes are related to the breaking of $O(3) \times O(2)$ 
symmetry by $\sqrt{3} \times \sqrt{3}$ ordering~\cite{DombreRead},  
and are: (i) a homogeneous 
 rotation of $A$ and $B$ spins, in phase with respect to upper and lower chains, and out of phase with respect to $A$ and $B$ directions, (ii) a homogeneous rotation of $A,~B$, and $C$ spins -- in phase with respect to upper and lower chains, and in phase with respect to $A$ and $B$, and $C$ spins,
  and (iii) a rotation with $k= \pi/2$,  in phase with respect to upper and lower chains, and in phase with respect to $A$ and $B$ spins, but  out of phase with respect to $C$ vs $A$ and $B$ spins. 

\begin{figure}[htb!]
\centering%
\includegraphics{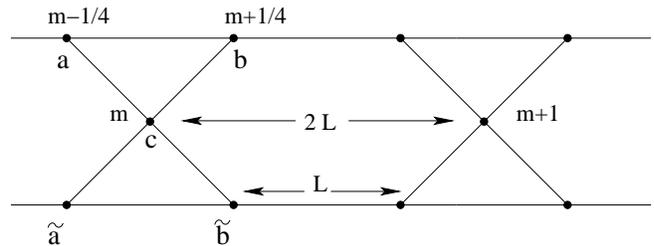} 
\caption{\label{fig:axeschoice1010}Notations for the Holstein-Primakoff and Fourier transformations. The distance between neighboring spins along the chains is  set to $L=1.$}
\end{figure}

We follow conventional strategy
 of the large-S approach: introduce five different bosonic operators 
 for five spins belonging to two top/bottom sharing triangles, 
 use Holstein-Primakoff
transformation from spin operators to bosons, and  
 diagonalize the quadratic form in boson operators. 
This is straightforward, but quite cumbersome procedure. 
We label the five bosons as shown in Fig. \ref{fig:axeschoice1010}, set 
 the distance between the sites along the
chains to be one, and introduce Fourier components as~\cite{KhveshchenkoChubukov87} 
$ a_{m-1/4}= \sum_k e^{2i k m} a_k, $ 
$ b_{m+1/4}= \sum_k e^{2i k m} b_k,$
$ c_{m}= \sum_k e^{2i k m} c_k,$ 
$ {\tilde a}_{m-1/4}= \sum_k e^{2i k m} {\tilde a}_{k}, $ 
$ {\tilde b}_{m+1/4}= \sum_k e^{2i k m} {\tilde b}_{k}.$
Substituting into Hamiltonian, we obtain $H = H_0 + H_2 + H_{int}$, where $H_0 = O(S^2)$ is the classical ground state energy, $H_2 = O(S)$ describes non-interacting spin-waves, and $H_{int} = O(1)$, which we neglect below, describes the interaction between spin-waves.   

For the spin-wave part, we obtain, explicitly 
\begin{multline}
H_2 =  J_1 S \sum_k  \left[ 
2 a_k^\dagger a_k +2 c_k^\dagger c_k+2 b_k^\dagger b_k+2 \tilde{a}_{k}^\dagger \tilde{a}_{k}  \right. \\
+ 2 \tilde{b}_{k}^\dagger \tilde{b}_{k} +\frac{1}{4} \left( a_k^\dagger b_k +a_k^\dagger c_k + b_k^\dagger c_k  \right.\\
\left.+  \tilde{a}_{k}^\dagger \tilde{b}_{k}+\tilde{a}_{k}^\dagger c_k + \tilde{b}_{k}^\dagger c_k + h.c. \right) \\ 
- \frac{3}{4} \left( a_k b_{-k} +a_k c_{-k}+b_k c_{-k} +\tilde{a}_{k}  \tilde{b}_{-k} \right. \\
\left.+\tilde{a}_{k} c_{-k} + \tilde{b}_{k} c_{-k} +  h.c. \right) \\
 \left.  - \left( e^{-i 2 k} b_k a_{-k} + e^{- i 2 k}  \tilde{b}_{k} \tilde{a}_{-k} + h.c. \right) \right]  \\
+ J_2 S\sum_k \left[a_k^\dagger a_k+b_k^\dagger b_k+ \tilde{a}_{k}^\dagger \tilde{a}_{k} +  \tilde{b}_{k}^\dagger \tilde{b}_{k} \right. \\
 \left.- \left( a^\dagger_k \tilde{a}_{k} + b^\dagger_k \tilde{b}_{k}  +h.c.\right) \right] \\
+ 2 J_3 S\sum_k \left[c^\dagger_k c_k - \frac{\cos(2 k)}{2} \left( c_k c_{-k} + h.c. \right)\right].
\label{eq:H2}
\end{multline}

Introducing the combinations of operators
$$
a_{1,2}(k) = \frac{1}{\sqrt{2}} ( a_{k} \pm \tilde{a}_{k} ), 
b_{1,2}(k)= \frac{1}{\sqrt{2}} ( b_{k} \pm  \tilde{b}_{k} ), 
$$
we find that the quadratic form is decoupled into a quadratic form which involves $a_1(k), ~b_1(k)$, and $c_{k}$ operators (which describe in-phase rotations of the spins in the upper and lower chains), and the quadratic form which involves only 
$a_2(k)$ and $b_2(k)$ operators (it describes out-of-phase rotations of the spins in the upper and lower chains). We then have 
$$H_2 = H_{2 \times 2} + H_{3 \times 3},$$
where
\begin{multline}
H_{2 \times 2} = J_1 S\sum_k \left\{
 2 a_2^\dagger(k) a_2(k) +2 b_2^\dagger(k) b_2(k) \right.\\ 
+ \frac{1}{4}  \left(a_2^\dagger(k) b_2(k) + h.c. \right)  \\
\left.-  \left[ \left(\frac{3}{4} + e^{i 2k} \right) a_2(k) b_2(-k) + h.c.  \right] \right\} \\
+J_2 S\sum_k  2 a_2^\dagger(k) a_2(k) +2 b_2^\dagger(k) b_2(k),
\label{eq:H2by2}
\end{multline}
\begin{multline}
H_{3 \times 3} = J_1 S \sum_k \left\{2   a_1^\dagger(k) a_1(k) + 2 b_1^\dagger(k) b_1(k)  \right. \\
+ \frac{1}{4}  \left(a_1^\dagger(k) b_1(k) +h.c.\right) \\
 -  \left[ \left(\frac{3}{4}+ e^{i 2k} \right) a_1(k) b_1(-k) + h.c.  \right]  \\
+ 2   c_{k}^\dagger c_{k}+\frac{\sqrt{2}}{4}  \left[ A^\dagger_1(k) c_{k} +B^\dagger_1(k) c_{k} + h.c. \right] \\
\left.-\frac{3 \sqrt{2}}{4}  \left[ a_1(k)  c_{-k}  +b_1(k) c_{-k} + h.c. \right] \right\} \\
+ 2 J_3 S \sum_k \left[c_{k}^\dagger c_{k} - \frac{\cos (2k)}{2} \left( c_{k} c_{-k}  + c_{k}^\dagger c_{-k}^\dagger \right) \right].
\label{eq:H3by3}
\end{multline}
\begin{figure}[t!]
\centering%
\includegraphics{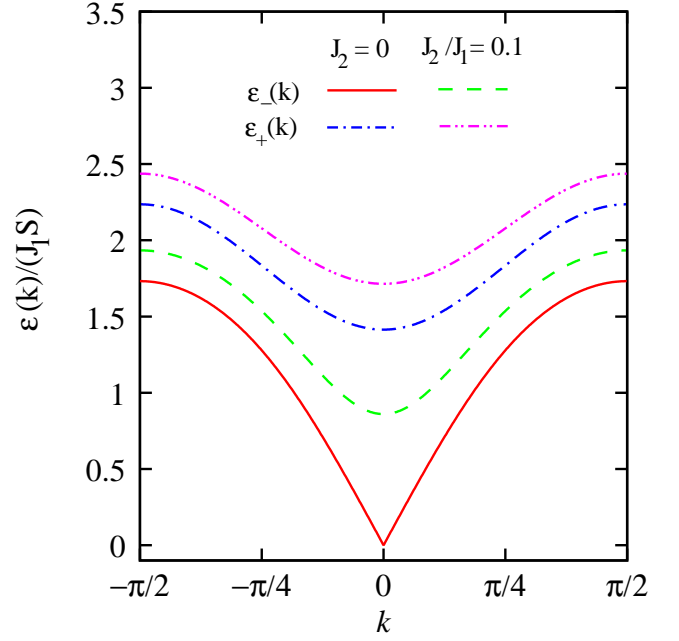}
\caption{\label{fig:Figure2by2} The spectrum of $H_{2 \times 2}.$ 
The two branches  $\epsilon_{\pm}$ are shown for $J_2=0$ 
(full line and dash-doted line), and for $J_2/J_1 = 0.1$ (dashed and dashed-two-dotted lines). A finite $J_2$ removes the zero mode at $k=0$. }
\end{figure}
The $H_{2 \times 2}$  part
depends on $J_1$ and $J_2$ but not on $J_3$.  Like we said, it describes 
 out-of-phase rotations of the spins in the upper and lower chains. Such rotations are not Goldstone modes, and generally the two spin-wave branches of  $H_{2 \times 2}$  must be gapped. This, however, happens only at a finite $J_2$; 
 for only neatest-neighbor interaction, out-of-phase rotations give rise to an extra zero mode (see Fig.~\ref{fig:J2lift}). 
 
The Hamiltonian  $H_{2 \times 2}$  can be diagonalized analytically. For this, we note that for a generic Hamiltonian 
\begin{multline}
 H_{4 \times 4} = \sum_k C_1 (k) \left( A_k^\dagger A_k + B_k^\dagger B_k \right) + C_2 (k)  \left(A_k^\dagger B_k \right. \\
\left.+B_k^\dagger A_k\right) + \left( C_3 (k)  A_k B_{-k} + C^*_3 (k) A_k^\dagger B_{-k}^\dagger \right),
\end{multline}
with  real $ C_1, C_2$ and complex $C_3$, the excitation spectrum is 
\begin{equation}
\varepsilon^2_{\pm} (k)  =
 \left[ \sqrt{C^2_1 (k) - (\text{Im}C_3 (k))^2} \pm C_2 (k) \right]^2 - (\text{Re}C_3 (k))^2.\label{n_1}
\end{equation}
 In our case, we have $C_1 (k) = 2 S (J_1 + J_2), \;\;C_2 (k) = J_1 S /4, \;\;
C_3 (k)  = - J_1 S \left( \frac{3}{4} + e^{2 i k} \right)$, 
such that $ \text{Re}  C_3 (k) = J_1 S \left( -3/4 - \cos (2k) \right), \;\;
\text{Im} C_3 (k)  = - J_1 S \sin (2k)$. Substituting into (\ref{n_1}), we obtain 
 two branches of excitations with the dispersion:
\begin{multline}
\varepsilon_{\pm} (k) = J_1 S \left\{\frac{5}{2} - \frac{3 \cos (2k)}{2} + \left( \frac{8 J_2}{J_1}+\frac{4 J_2^2}{J_1^2}\right) \right. \\
\pm \frac{1}{2} \left. \left[ \frac{7}{2}+\frac{\cos (4k)}{2} + \left(\frac{8 J_2}{J_1}+\frac{4 J_2^2}{J_1^2} \right) \right]^{1/2} \right\}^{1/2}.
\end{multline}
The spectrum is shown in Fig.\ref{fig:Figure2by2} for $J_2=0$ and $J_2/J_1=0.1.$ In the absence of $J_2$, the dispersion $\varepsilon_{-} (k)$ has a zero mode at
 $k=0$. At a nonzero $J_2$, the zero mode is lifted, and the dispersion acquires a finite gap, as we anticipated. The other dispersion branch, 
$\varepsilon_{+} (k)$, is gapped already at $J_2=0$, and does not change significantly with $J_2$. 

\begin{figure}[tb!]
\centering%
\includegraphics{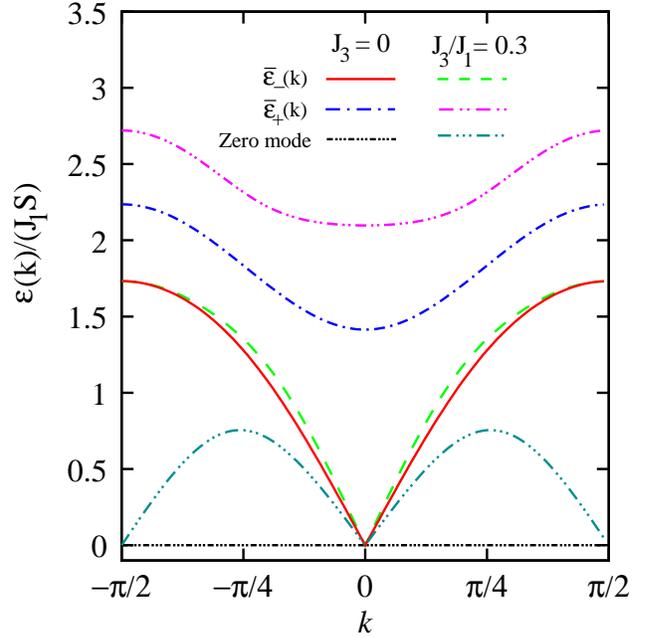}
\caption{The spectrum of $H_{3 \times 3}.$  Without $J_3$, the dispersion has 
one linear in $k$ (Goldstone) branch, and one branch of zero energy excitations. When $J_3$
 is finite, the former zero-energy branch acquires a dispersion and 
 yields two Goldstone modes at $k=0$ and $k = \pi/2$. The third branch is gapped with and without $J_3$.}
\label{fig:Figure3by3}
\end{figure}

The diagonalization of the $H_{3\times3}$ part of the Hamiltonian is more involved as this Hamiltonian contains three operators with complex coefficients. 
The diagonalization amounts to solving $6$ by $6$ matrix which we did numerically. The results for $J_3 =0$ and  $J_3=0.3 J_1$  are plotted in Fig.~\ref{fig:Figure3by3} 
There are three branches of magnon dispersion. They describe 
in phase and out-of-phase rotations between $A$, $B$, and $C$ spins -- in all cases the rotations are symmetric with respect to spins in the upper and lower chains. The three Goldstone modes are among these excitations. 

For $J_3=0$, one of the branches is gapped, another is linear in $k$ near $k=0$, and the third one is exactly zero for all $k-$points.  This is the consequence of the local degeneracy. All three
 branches  can be easily obtained analytically as at $J_3 =0$, the 
 zero-energy branch decouples from the other two branches, which are 
 the same as $\varepsilon_{\pm} (k)$ for $J_2=0$. At a 
 finite $J_3$, the local degeneracy is lifted, and the former zero-energy branch acquires a  ``$\sin 2k$''-like dispersion with finite energy
 at a generic $k$, and Goldstone points at $k=0$ and $k= \pm \pi/2$. 
 The linear in $k$ branch remains gapless
 and its velocity is only slightly affected by $J_3$.  

The existence of three Goldstone modes at a finite  $J_3$ 
agrees with what one should expect on general grounds.  Furthermore, the 
 Goldstone modes can be obtained analytically. For $k=0$ and $k=\pi/2$, the 
 out-of-phase mode $p_- = (a_1-b_1)/\sqrt{2}$ is decoupled from the 
  modes $p_+ = (a_1 + b_1)/\sqrt{2}$ and $c$. At k=0, the energy of $p_{-}$ mode
 is zero (Goldstone), at $k =\pi/2$, it is $J_1 S \sqrt{3}$, independent on $J_3$.  For the remaining two coupled excitations, $p_+$ and $c$, the Hamiltonian can be diagonalized in the same spirit as $H_{2\times 2}$. At $k=0$, we obtain
 one solution at zero energy (Goldstone), and one at energy ${\bar \varepsilon}_{+} (k=0) = J_1 S \sqrt{2 + 8 J_3/J_1}$, at
 $k=\pi/2$, one solution has zero energy (Goldstone), another is at energy
${\bar \varepsilon}_{+} (\pi/2) = J_1 S \sqrt{5 + 8 J_3/J_1}$. The 
 combinations of $p_{+}$ and $c$ for the Goldstone modes are $2p_{+} + c$ at $k=0$
 and $p_{+} -c$ at $k =\pi/2$. 

In summary, in this paper, we considered  a Heisenberg model on a
 Kagome stripe.  We showed that for nearest-neighbor interaction, the quasiclassical excitation spectrum contains two special  features associated with the 
 extra degeneracies of a classical ground state,
  a zero-energy branch  of excitations associated with a local degeneracy,
 and a single  zero mode associated with an additional
 global degeneracy. We demonstrated that the 
 zero mode and zero-energy branch are removed by second- and third--neighbor
 interactions, respectively. We obtained the full quasiclassical spin-wave spectrum for a model with the interaction between first, second, and third neighbors, and showed that it contains three Goldstone modes, precisely as it should 
 be for a ground state which breaks $O(3) \times O(2)$ symmetry.

A.V.C. acknowledge the support from
NSF-DMR 0604406, and the hospitality of the TU Braunschweig and MPIPKS in Dresden during the completion of this work.

\end{document}